\documentclass[a4paper,twoside,openany,11pt]{memoir} 

\def\fullheadfoot{0}
\usepackage[english]{babel}
\usepackage{microtype,xspace}
\usepackage[utf8]{inputenc}	

\usepackage{amsfonts,amsmath,amssymb,bm,mathrsfs,mathtools,dsfont} 	
\allowdisplaybreaks 	
\usepackage{xcolor}
\usepackage{hyperref}
\usepackage{slashed}
\usepackage{multicol}

\usepackage{siunitx}
\usepackage{geometry}
\usepackage[sort&compress,numbers,merge]{natbib}
\usepackage{chapterbib}
\addto\captionsenglish{\renewcommand*{\bibname}{References}}
\makeatletter
\renewcommand{\@memb@bchap}{ 
\bibmark \prebibhook
}
\makeatother

\usepackage{graphicx}
\graphicspath{{./Figures/}}
\usepackage{caption,subcaption}
\usepackage{multirow}
\renewcommand{\arraystretch}{1.2}
\usepackage{tabularx}
\newcolumntype{Y}{>{\centering\arraybackslash}X}

\usepackage[shortlabels]{enumitem}
\setlist{itemsep=.1em,topsep=.5em}
\SetEnumerateShortLabel{i}{\textit{\roman*}}

\definecolor{red}{rgb}{0.6,.0706,.1373}
\definecolor{blue}{rgb}{0,0.396,0.741}
\definecolor{green}{rgb}{0.3,0.55,0.2}
\definecolor{yellow}{rgb}{0.7,0.63,0.062}
\colorlet{blueRef}{blue!80!black}
\colorlet{blueLink}{blue!100!black}
\hypersetup{
	colorlinks, 
	bookmarksopen, 
	bookmarksnumbered,
	citecolor=blueLink, 	
	linkcolor=blueLink,	
	urlcolor=blueLink,			
	pdftitle={Stability of the Higgs Sector},    
	pdfsubject={BSM physics},	
	pdfauthor={L. Allwicher, G. Isidori, and A.E. Thomsen},    	
}

\addto\captionsenglish{}

\renewcommand{\contentsname}{Contents}
\renewcommand{\printtoctitle}[1]{}

\settocdepth{subsection}
\makeatletter
\newcommand*\ifthispageodd{%
  \checkoddpage
  \ifoddpage
    \expandafter\@firstoftwo
  \else
    \expandafter\@secondoftwo
  \fi
}
\makeatother

\usepackage[noindentafter]{titlesec} 
\counterwithout{section}{chapter} 
\counterwithout{figure}{chapter} 
\counterwithout{table}{chapter} 
\setsecnumdepth{subsubsection}

\DeclareMathAlphabet{\mathsfit}{OT1}{lmss}{m}{sl}
\DeclareMathAlphabet{\mathsfbf}{OT1}{lmss}{bx}{n}
\DeclareMathAlphabet{\mathsfbfit}{OT1}{lmss}{bx}{sl}
 
\DeclareMathVersion{chaptermath}
\SetSymbolFont{operators}{chaptermath}{OT1}{cmbr}{m}{n}
\SetSymbolFont{letters}{chaptermath}{OML}{cmbrm}{b}{it}
\SetSymbolFont{symbols}{chaptermath}{OMS}{cmbrs}{m}{n}
\DeclareMathVersion{subsectionmath}
\SetSymbolFont{operators}{subsectionmath}{OT1}{cmbr}{m}{n}
\SetSymbolFont{letters}{subsectionmath}{OML}{cmbrm}{m}{it}
\SetSymbolFont{symbols}{subsectionmath}{OMS}{cmbrs}{m}{n}

\titleformat{\section}{\centering \Large \bfseries \sffamily \mathversion{chaptermath} \color{blue!90!black} }{\thesection}{15pt}{}{}
\titlespacing{\section}{0pt}{15pt}{5pt}
\titleformat{\subsection}{\large \sffamily \mathversion{subsectionmath} \color{blue!90!black} }{\thesubsection}{10pt}{}{}
\titlespacing{\subsection}{0pt}{10pt}{5pt}
\titleformat{\subsubsection}{\normalsize \sffamily \itshape \mathversion{subsectionmath} \color{blue!80!black} }{\thesubsubsection}{10pt}{}{}
\titlespacing{\subsubsection}{0pt}{10pt}{0pt}

\newcommand{\sectionlike}[1]{\phantomsection \addcontentsline{toc}{section}{#1} \sectionmark{#1}
		\begin{center}
		\needspace{8\baselineskip}
		\Large \bfseries \sffamily \mathversion{chaptermath} \color{blue!90!black} #1  
		\end{center}
	\vspace{-5pt} 
}

\ifnum\fullheadfoot=1
	\makepagestyle{standardstyle}
	\makeoddhead{standardstyle}{\sffamily \mathversion{subsectionmath} \rightmark}{}{\sffamily \bfseries{\thepage}}
	\makeevenhead{standardstyle}{\sffamily \bfseries{\thepage}}{}{\sffamily \mathversion{subsectionmath} \leftmark}
	\makeheadrule{standardstyle}{\textwidth}{\normalrulethickness}
	\makepsmarks{standardstyle}{
		\nouppercaseheads
		\createmark{section}{both}{shownumber}{\ }{\hspace{1.2em}  }
		\createmark{subsection}{right}{shownumber}{}{\hspace{1.2em} }
	}
	
	\copypagestyle{appstyle}{standardstyle}
	\makeevenhead{appstyle}{\sffamily \bfseries{\thepage}}{}{ \sffamily \mathversion{subsectionmath} \leftmark}
	\makepsmarks{appstyle}{
		\nouppercaseheads
		\createmark{section}{both}{nonumber}{\ }{\ \ }
		\createmark{subsection}{right}{shownumber}{}{\ \ }
	}
\else 
	\makepagestyle{standardstyle}
	\makeoddfoot{standardstyle}{}{\sffamily -- {\bfseries\thepage} --}{} 
	\makeevenfoot{standardstyle}{}{\sffamily -- {\bfseries\thepage} --}{}
	\copypagestyle{appstyle}{standardstyle}
\fi

\createplainmark{toc}{both}{\contentsname}
\createplainmark{bib}{both}{\bibname}

\makeatletter
\let\MyIntOrig\int
\def\MyIntSpace{\hspace{-.35em}} 
\def\int{\MyInt}
\def\MyInt{\MyIntOrig\MyIntSkipMaybe}
\def\MyIntSkipMaybe{
	\@ifnextchar_{\MyIntSkipScript}{%
		\@ifnextchar^{\MyIntSkipScript}{%
			\@ifnextchar\limits{\MyIntSkipTok}{%
				\@ifnextchar\nolimits{\MyIntSkipTok}{%
					\MyIntSpace}}}}%
}
\def\MyIntSkipScript#1#2{#1{#2}\MyIntSkipMaybe}
\def\MyIntSkipTok#1{#1\MyIntSkipMaybe}

\newcommand{\pushright}[1]{\ifmeasuring@#1\else\omit\hfill$\displaystyle#1$\fi\ignorespaces}
\makeatother

\newcommand{\brakets}[1]{\big\langle #1 \big\rangle}
\newcommand{\Tr}[1]{\mathop{\mathrm{Tr}}\!\left[#1\right]\!}

\newcommand{\eminus}{\vcenter{\hbox{\scalebox{0.6}[1]{$ - $}}}}	
\newcommand{\ord}[1]{\mathcal{O}( #1 )}

\newcommand{\hc}{\; + \; \mathrm{H.c.} \;}

\newcommand{\diag}{\mathop{\mathrm{diag}}}
\newcommand{\transpose}{^{\mathrm{T}}}
\newcommand{\rep}[1]{\mathbf{#1}}
\newcommand{\repbar}[1]{\overline{\mathbf{#1}}}

\newcommand{\ineqgraphics}[1]{\vcenter{\hbox{\includegraphics[]{#1}}}}

\makeatletter
\newcommand{\vast}{\bBigg@{3}}
\makeatother

\newcommand{\be}{\begin{equation}}
\newcommand{\ee}{\end{equation}}
\newcommand{\bea}{\begin{eqnarray}}
\newcommand{\eea}{\end{eqnarray}}
\newcommand{\cL}{\mathcal{L}}
\newcommand{\cO}{\mathcal{O}}
\newcommand{\mut}{\tilde \mu}
\newcommand{\Ht}{\tilde H}
\newcommand{\vv}{v_\mathrm{EW}}

\renewcommand{\L}{\mathcal{L}}
\newcommand{\LL}{\mathrm{L}}
\newcommand{\YY}{\mathrm{Y}}
\newcommand{\RR}{\mathrm{R}}
\newcommand{\U}{\mathrm{U}}
\newcommand{\SU}{\mathrm{SU}}

\newcommand{\SO}{\mathrm{SO}}

\begin{document}

\thispagestyle{empty}
\renewcommand*{\thefootnote}{\fnsymbol{footnote}}
\vspace*{0.05\textheight}
\begin{center}
	{\sffamily \bfseries \huge \mathversion{chaptermath} 
	Stability of the Higgs Sector\\[2pt]  in a Flavor-Inspired Multi-Scale Model}\\[-.5em]
	\textcolor{blue!80!black}{\rule{.9\textwidth}{2.5pt}}\\
	\vspace{.02\textheight}
	{\sffamily \Large Lukas Allwicher,\footnote{lukas.allwicher@physik.uzh.ch} 
	Gino Isidori,\footnote{isidori@physik.uzh.ch} 
	and Anders Eller Thomsen\footnote{thomsen@itp.unibe.ch}}
	\\{\normalsize \sffamily 
	Physik-Institut, Universit\"at Z\"urich, CH-8057 Z\"urich, Switzerland
	}
	\\[.005\textheight]{\itshape \sffamily \today}
	\\[.03\textheight]
\end{center}
\setcounter{footnote}{0}
\renewcommand*{\thefootnote}{\arabic{footnote}}%
\suppressfloats	

\begin{abstract}\vspace{-.03\textheight}
We analyze the stability of the Higgs sector of a three-site model with flavor-non-universal gauge interactions, whose spectrum 
of non-Standard-Model states spans three orders of magnitude. This model is inspired by  deconstructing  
a  five-dimensional theory where the generation index is in one-to-one relation to the position in the fifth dimension.
It provides a good description of masses and mixing of the SM fermions in terms of scale hierarchies. 
We demonstrate that, within this construction, the mass term of the SM-like Higgs 
does not receive large corrections proportional to the highest mass scales. 
The model suffers only of the unavoidable ``little hierarchy problem'' between 
the electroweak scale and the lightest NP states, which are expected to be at the TeV scale. 

{\vspace{.7em} \footnotesize \itshape Preprint: ZU-TH-41/20}
\end{abstract}

\section{Introduction}
Trying to stabilize the Standard Model (SM) Higgs sector against quantum corrections is a central theme in particle physics.
The quadratic sensitivity of the Higgs mass term to ultraviolet degrees of freedom suggests the existence of new states 
around the TeV scale able to ``screen''  the SM ground state from possible higher scales.  That we have not yet 
seen any new particle at the LHC experiments is posing a serious challenge to this general argument~\cite{giudice2013naturalness}.
However, rather than completely abandoning the naturalness criterion~\cite{tHooft:1980xss} in this context, 
it is worth further investigating how serious the present challenge is.

A quantification of the tuning necessary to stabilize the Higgs mass in a given Beyond SM (BSM) 
framework is a model-dependent question  that can be addressed only after specifying how heavy,
and how strongly coupled to the SM fields,  the new degrees of freedom are. Several analyses of this 
type exist in the literature for specific mechanisms focused on the stabilization of the Higgs sector
(or, more precisely, the stabilization of the Higgs--gauge--top sector).
In this paper we would like to address a  different but closely related question, 
namely how addressing the flavor problem close to the electroweak scale 
affects the stability of the Higgs sector. 

When considering the SM as a low-energy effective theory, we encounter a twofold issue 
in the flavor sector: there is no explanation for the hierarchical strucutre of the Yukawa couplings, 
and there are very severe bounds on possible non-SM sources of flavor violation~\cite{Isidori:2010kg}. We denote the 
combination of these two issues the flavor problem. In a wide class of BSM frameworks,
the flavor problem and the stabilization of the Higgs sector are dealt with separately. 
The paradigmatic example is models based on the Minimal Flavor Violation (MFV) 
hypothesis~\cite{Chivukula:1987py,DAmbrosio:2002vsn}.
Within this class of models, the new degrees of freedom stabilizing the Higgs sector are assumed to be flavor blind, 
but for correction terms proportional to the SM Yukawa couplings. 
While minimizing New Physics (NP) constraints from low-energy flavor-violating observables, 
nowadays 
the MFV hypothesis worsens the electroweak hierarchy problem, given that high-$p_T$ bounds are 
particularly strong for NP coupled in a family-universal way. 
 
The opposite point of view is that of flavor-non-universal interactions already at the TeV scale. 
This approach is motivated by the smallness of light fermion masses: if we assume that the 
Yukawa interactions are the low-energy effect of some new dynamics, it is natural to consider 
a NP framework with a hierarchical spectrum with TeV scale dynamics
coupled only to third generation fermions and heavier states coupled to the light SM fermions.   
A partial realization of this mechanism occurs in models with partial 
compositeness~\cite{Kaplan:1991dc,Gherghetta:2000qt,Contino:2003ve,Agashe:2004cp}.
However, in that context gauge interactions and, to a large extent,  the dynamics stabilizing the Higgs sector, 
are still flavor universal. A more radical approach 
is that of flavor-non-universal gauge interactions, such as the $\textrm{PS}^3$ framework proposed in~\cite{Bordone:2017bld}. 
The latter model, which is phenomenologically motivated by the recent $B$-physics anomalies
(see e.g.~\cite{deSimone:2020kwi} and references therein),
is a prototype of the BSM framework we are interested in. 
On general grounds, this class of models has a clear advantage in escaping the NP bounds from 
direct searches. However, it might suffer from a more severe tuning in the Higgs sector due to the 
heavy states coupled to the light generations, which are charged under the SM gauge group.
This potential problem is the issue we would like to investigate in this paper.
 
The underlying idea behind flavor-non-universal gauge interactions is that what we denote as ``flavor''  or ``generation'' is related to a specific localization 
in an extra space-like dimension. More precisely, following the  principle of dimensional (de)construction~\cite{ArkaniHamed:2001ca}, 
we can establish a one-to-one relation between the four-dimensional (4D) model and a five-dimensional 
(5D) theory where the fifth dimension is discretized, and the three flavor indices denote three notable points
along this discretized dimension. For a more explicit construction of this type see e.g. Ref.~\cite{Fujimoto:2012wv}. 
The 5D picture is  instructive to provide a rational explanation for the  
peculiar structure of fields and couplings of the model, while the 4D description is more convenient
in order to precisely compute amplitudes beyond the tree level. 

In order to investigate the  stability of the Higgs sector, in general terms,  in this class of models,
we employ the following strategy. In Section~2 we briefly outline the rules on how to build 
renormalizable 4D models based on the idea of an underlying 5D theory, where flavor denotes the location 
in the fifth dimension. We then proceed checking explicitly how the scale mixes under quantum corrections
in these 4D constructions. 
In Section~3 we start analyzing a simplified model able to address the fermion mass hierarchies (relating them to different underlying scales), but not the fermion mixing. This allow us to introduce and discuss some general aspects of the construction and the stability of the Higgs-gauge sector.  The issue of fermion mixing and the related stability of the Yukawa sector is discussed in Section~4 within 
the context of a more realistic model.
The results are summarized in the Conclusions.

\section{General considerations}
The general structure of the models we are interested in is a chain of identical gauge groups,
labeled by a  flavor (or generation) index $i$, with each fermion family charged only under one of these gauge groups.
This strucutre is inspired by the deconstruction in ordinary four-dimensional (4D) space time 
of models defined in $d=4+1$, where the fourth spacial dimension is compact and discretized into three lattice
points corresponding to two fundamental intervals, or two fundamental scales. These points 
are   in one-to-one correspondence with the flavor index. 
The gauge symmetry of the 5D theory reduces 
to a product of three identical 4D gauge groups, and a series of scalar (link) fields, $\Omega_{i(i+1)} $, 
transforming as bilinears of the (4D) gauge symmetry~\cite{Hill:2000mu,Cheng:2001vd}. 
The vacuum expectation value (VEV) of the link fields breaks the chain of gauge groups to the diagonal 
(flavor-universal) subgroup, which we identify with the electroweak SM gauge group, or at least with part of it.

To this underlying gauge and fermion structure, we add a series of Higgs fields,
characterized by the flavor index $i$, which can be interpreted as the remnant of a unique 5D Higgs field with
non-trivial dependence on the discretized dimension. 
We will not venture into the details of the 5D construction beyond the key point that tree-level interactions between different sites in the class of 4D models under consideration only occur via insertions of appropriate link fields. From the 5D perspective, this assumption can be viewed as the effect of fermion fields being localized on the three lattice points. The Higgs field responsible for the 
electroweak symmetry breaking (EWSB) 
is transmitted to the other sites through a slight mass mixing with the other Higgs states. Generically, this requires a hierarchical mass structure, which is reflected into an inverse hierarchy for the Higgs-Yukawa couplings thereby addressing the flavor problem. Similar constructions have also been pursued in fundamental 4D theories~\cite{Porto:2007ed,Giudice:2008uua,Hill:2019ldq}.
 
Our goal is to study the stability of the Higgs potential in the presence of such a mass hierarchy. 
For this reason, both in the simplified model analyzed in Section~\ref{sect:simp_model}
and in the more realistic model discussed in Section~\ref{sec:full_model}, 
we limit ourselves to consider a chain of $\SU(2)$ groups, with the diagonal  $\SU(2)_{1+2+3}$ identified with the 
$\SU(2)_\LL $ gauge group of the SM. The rest of the SM gauge sector, which is not particularly relevant to our discussion,
is treated as a flavor-universal group.

\section{Simplified three-site model} \label{sect:simp_model}
The matter field content (fermions and scalars) of the simplified model is summarized in Tab.~\ref{tab:min_model}. 
For simplicity, we consider only up-type quark singlets, though down-type can be added analogously. The Yukawa couplings for such fields, defined according to the 
rules specified above are 
	\be \label{eq:Higgs_Yukawa_couplings}
	\cL_Y = - \sum_{i=1}^{3} y_i \overline{q}_{i,\LL} H^{\ast}_i u_{i,\RR} \hc
	\ee
Assigning (flavor-universal) $\U(1)_\YY$ charges, as in the SM, to fermions and Higgs fields forbids 
any additional Yukawa interactions among the fields listed in Tab.~\ref{tab:min_model}. 
The SM-like Higgs field, $ \Ht_3 $, is the combination of the three $H_i$ fields that gets a non-vanishing VEV. Thanks to  $\cL_Y$ of Eq.~\eqref{eq:Higgs_Yukawa_couplings}, this VEV gives rise to non-vanishing masses for all three fermion generations, 
whose values are controlled by $y_i$, assumed to be of $\cO(1)$, and the Higgs mixing terms, which exhibit a hierarchical structure.
	
\begin{table}
	\centering
	\begin{tabular}{| c | c c|}
	\hline\hline
	Fields & $ \SO(1,3) $ & $ [\SU(2)]^3 $ \\
	\hline
	$ q_i $ & $ (\tfrac{1}{2},\, 0) $ & $ \rep{2}_i $ \\
	$ u_i $ & $ (0,\, \tfrac{1}{2}) $ & $ \rep{1} $ \\
	\hline
	$ H_i $ & 0 & $ \rep{2}_i $  \\
	$ \Omega_{i(i+1)} $ & 0 & $ \rep{2}_i \otimes \repbar{2}_{i+1} $ \\
	\hline \hline
	\end{tabular}
	\caption{Matter field content of the minimal three-site model.}
	\label{tab:min_model}
\end{table}

\subsection{Higgs sector}

The Higgs mixing is caused by the coupling of the $ H_i $'s to the link fields, $ \Omega_{ij} $, whose VEVs break the gauge symmetry down to the diagonal subgroup. We will not go into the details of the link field potential but merely assume it to be such that they develop VEVs:
\be
\brakets{\Omega_{i(i+1)}} = F_{i(i+1)}~,   \qquad  \textrm{with} \qquad  F_{12} \gg F_{23} \gg \vv~,
\ee
where $\vv$ denotes the  VEV of the SM-like Higgs.
The relevant scalar couplings involving Higgs and link fields are 
	\bea
	\L & \supset & - \sum_{i=1}^{3} m_i^2 H_i^\dagger H_i + \bigg(\sum_{i=1}^{2} \kappa_{i(i+1)}\,  F_{i(i+1)} H_i^\dagger \Omega_{i(i+1)} H_{i+1} \hc \bigg) \nonumber \\
	&& 	- \sum_{i=1}^2  \lambda^+_{i}\, H_i^\dagger \Omega_{i(i+1)}\Omega_{i(i+1)}^\dagger H_i 
   	- \sum_{i=2}^3 \lambda^-_{i}\, H_i^\dagger \Omega_{(i-1)i}^\dagger\Omega_{(i-1)i}H_i~, \label{eq:LHiggs1}
	\eea
where $\kappa_{i(i+1)}$ and $\lambda^\pm_{i}$ are adimensional 
couplings. In a 5D description,  all the interaction terms above can be derived starting from the 5D kinetic term of the unique 
Higgs field with non-trivial profile in the fifth dimension. However, the values of the couplings can be modified by adding 
non-trivial boundary conditions on the discrete points, which is why in the following we treat 
$\kappa_{i(i+1)}$ and $\lambda^\pm_{i}$ as free parameters. As we shall see, the consistency 
of the construction requires $ \lambda_{2,3}^{-} \ll 1$, while all the other couplings can be of $\cO(1)$.

From a pure 4D point of view, additional quartic couplings involving $ \Omega_{ij} $ and $ H_i $ are possible.
However, our goal is to study the quadratic part of the effective Higgs potential and such terms do not contribute 
to it at the tree level, contrary to the terms in Eq.~\eqref{eq:LHiggs1}, 
which do provide a contribution once the link fields acquire VEVs.\footnote{From the
underlying 5D construction, we expect quartic couplings involving only one type of Higgs field to be present at the tree level, 
whereas mixed terms of the type $ (H_i^{\dagger}  H_i^{\dagger})(H_j^{\dagger}  H_j^{\dagger})$ to be radiatively generated
and to appear in the potential with small couplings. In principle, the  operator  $H_1^\dagger \Omega_{12} \Omega_{23} H_3 $  would 
modify  the quadratic part of the Higgs potential already at the tree level; however, employing 5D arguments, we 
can assume the corresponding coupling to be very suppressed.}
To take into account the modification of the quadratic  terms for $\brakets{\Omega_{i(i+1)}} = F_{i(i+1)}$, 
it is convenient to redefine the  $m_i^{2}$ and treat them as the effective tree-level Higgs masses 
after the link fields acquire VEVs. The stability of the potential under quantum 
corrections is discussed in the next subsection.

The key assumption on the various scales of the system (the link VEVs, $F_{ij}$, and the flavor-diagonal effective Higgs masses, $m_i$), which we employ in order to ensure both a correct hierarchy for the Yukawa couplings and 
a stable system under quantum corrections, is 
\be
	m_1  \gg \; F_{12} \sim \varepsilon\, m_1 \; \gg \; m_2 \sim \varepsilon^2\, m_1 \; \gg \; F_{23} \sim \varepsilon^3\, m_1	
	\; \gg \; m_3 \sim  \vv \sim \varepsilon^4\, m_1~,
\label{eq:scaleratios}
\ee
where $\varepsilon$ is  a small parameter of $\ord{10^{\eminus 1}}$. As we shall see,  $\varepsilon  \sim  (4\pi)^{\eminus 1} $
is the smallest value for such parameter compatible with radiative stability. 
For $ \lambda_{2,3}^{-} = \cO(1)$,
the scales in Eq. \eqref{eq:scaleratios} implies an $\cO(\varepsilon^2)$ tuning between the  $m^2_i$ before 
the link fields acquire VEVs, and the effective  $m^2_i$ after the $ [\SU(2)]^3  \to \SU(2)_{1+2+3}$
symmetry breaking, for both $i=2$ and $i=3$.
This is nothing but a manifestation of the  little hierarchy problem~\cite{giudice2013naturalness} in this context. 
This tuning can be 
ameliorated assuming $|\lambda^-_{2,3}|\ll1$, which in practice is equivalent to shifting the tuning  
to a different set of couplings. We will simply accept this tuning as a specific relation among 
couplings at low energies, as we address the question of stability under quantum corrections. 

The hierarchy in Eq.~\eqref{eq:scaleratios} allows us to perform a perturbative diagonalization of the $ 3\times3$ quadratic 
piece of the Higgs potential, namely 
	\begin{equation}
	\mu_H^2  = 
	\begin{pmatrix}
	m_1^2 & \kappa_{12} F_{12}^2   & 0 \\ 
	\kappa_{12} F_{12}^2  & m_2^2 & \kappa_{23} F_{23}^2  \\
	0 & \kappa_{23} F_{23}^2   & m_3^2 
	\end{pmatrix} \sim 
	m_1^2 
	\begin{pmatrix}  
	1 &  \varepsilon^2  &  0 \\   \varepsilon^2  & \varepsilon^4 & \varepsilon^6 \\  0 & \varepsilon^6 & \varepsilon^8 
	\end{pmatrix}~.
	\end{equation}
Defining 
\be
\mu_H^2  = O_H\,   \textrm{diag}( \mut_1^2, \mut_2^2 , \mut_3^2 )\,  O_H\transpose \equiv  O_H\, \mut_H^2\, O_H\transpose
\ee
and keeping each entry to leading order (LO) in the $ \varepsilon $ expansion, we obtain  
	\begin{equation} 
	\label{eq:OH}
	\renewcommand\arraystretch{2}
	O_H = \begin{pmatrix}
	1 & - \kappa_{12} \frac{F_{12}^2 }{m_1^2}   & \frac{\kappa_{12} \kappa_{23} F_{12}^2 F_{23}^2}{m_1^2 m_2^2 - \kappa_{12}^2 F_{12}^4}  \\
	\kappa_{12} \frac{F_{12}^2 }{m_1^2}  & 1 & -\frac{ \kappa_{23} F_{23}^2 m_1^2}{m_1^2 m_2^2 - \kappa_{12}^2 F_{12}^4} \\
	\frac{\kappa_{12} \kappa_{23} F_{12}^2 F_{23}^2}{m_1^4 }  & \frac{ \kappa_{23} F_{23}^2 m_1^2}{m_1^2 m_2^2 - \kappa_{12}^2 F_{12}^4}  & 1
	\end{pmatrix}
	\sim
	\begin{pmatrix}
	1 & \varepsilon^2  & \varepsilon^4 \\
	\varepsilon^2 & 1 & \varepsilon^2 \\
	\varepsilon^8 & \varepsilon^2 & 1
	\end{pmatrix}~,
	\end{equation}
and 
	\begin{equation}\label{eq:mut}
	\mut_H^2 = \diag\left(m_1^2, \  m_2^2 - \kappa_{12}^2\dfrac{F_{12}^4}{m_1^2},  \   m_3^2 - \kappa_{23}^2 \dfrac{ F_{23}^4 m_1^2}{m_1^2 m_2^2 - \kappa_{12}^2 F_{12}^4}  \right) \sim 
	 m^2_1\, \diag\left(1,\  \varepsilon^4, \  \varepsilon^8 \right)~.
	\end{equation}
In order to trigger an SM-like EWSB, we require $\mut^{2}_{1,2}  > 0 $ and $\mut^{2}_3 < 0$ with 
$|\mut_3| = \ord{100~\textrm{GeV}}$.
Since both terms in $\mut^{2}_3$ are of $\cO(\varepsilon^8)$, this requirement  
does not imply any additional tuning. 
With a suitable choice of the parameters controlling  quartic interactions,  
we can finally obtain  $\brakets{\Ht_3} = \vv$ for the 
lightest Higgs field, which we identify with the SM-like Higgs boson.

Returning to the flavor basis for the Higgs fields, it is easy to realize that the non-vanishing VEV of $\brakets{\Ht_3}$
is distributed to all the three fields in a hierarchical way. To leading non-vanishing order in the perturbative expansion, we get 
	\begin{equation}
	\label{eq:higgsVEVs}
	\brakets{H_1} = \dfrac{\kappa_{12} \kappa_{23} F_{12}^2 F_{23}^2}{m_1^2 m_2^2 - \kappa_{12}^2 F_{12}^4}  \binom{0}{v_\mathrm{EW}}, \qquad \brakets{H_2} = -\dfrac{ \kappa_{23} F_{23}^2 m_1^2}{m_1^2 m_2^2 - \kappa_{12}^2 F_{12}^4}  \binom{0}{v_\mathrm{EW}}, \qquad \brakets{H_3} = \binom{0}{v_\mathrm{EW}},
	\end{equation}
which, assuming $\cO(1)$ values for the Yukawa couplings~\eqref{eq:Higgs_Yukawa_couplings}, leads to up-quark mass matrix
	\begin{equation}
	m^{u}_{ij} \sim \diag(\varepsilon^4,\; \varepsilon^2,\; 1)\, v_\mathrm{EW}~.
	\end{equation}
The mechanism comfortably reproduces the observed hierarchy of the up-type quark masses with $ \varepsilon \sim (4\pi)^{\eminus 1}$. 

The attentive reader will have noticed the absence of any mixing in the fermion mass matrix. The latter is not well defined in absence of 
$\SU(2)$-singlet down-type quarks, but it is easy to verify that adding such fields and proceeding in a similar manner does not lead 
to any physical mixing. This is because the fermion part of the Lagrangian respects a $\U(1)^3$ global symmetry of the fermions on each site protecting against any such mixing. The more realistic model introduced in Sec.~\ref{sec:full_model} will deal with this issue by introducing heavy vector-like fermions and, with them, fermion mixing.

\subsection{Radiative stability of the Higgs-gauge sector} \label{sec:gauge_stability}
In order to discuss the stability of the Higgs potential under radiative corrections due to the gauge (and link) sector, 
we must first analyze the spectrum of the gauge fields in the model.
The latter is generated by the kinetic terms of the link fields
	\begin{equation}\label{eq:gauge_kin_term}
	\L \supset \sum_{i=1}^{2} \Tr{D_\mu \Omega_{i(i+1)}^{\dagger} D^\mu \Omega_{i(i+1)} }~,
	\end{equation}
which are responsible for the breaking $ [\SU(2)]^3 \to \SU(2)_{1+2+3} $.
Defining $ W_{i\mu} = W_{i\mu}^{a} \sigma^a/2 $, the covariant derivatives acting on the link fields are 
	\begin{equation}
	D_\mu \Omega_{i(i+1)} = \partial_{\mu} \Omega_{i(i+1)} - i g_i W_{i\mu} \Omega_{i(i+1)} +i g_{i+1} \Omega_{i(i+1)} W_{(i+1)\mu}. 
	\end{equation} 
The mass terms for the gauge fields are then quickly determined by expanding the covariant derivatives in Eq.~\eqref{eq:gauge_kin_term}: 	
	\begin{equation}
	\L \supset \dfrac{1}{2} W^{a}_{i\mu} (M_W^2)_{ij} W^{a\mu}_{j},\qquad M_W^2 =
	\begin{pmatrix} g_1^2 F_{12}^2 & g_1 g_2 F_{12}^2 & 0 \\ 
		g_1 g_2 F_{12}^2 & g_2^2(F_{12}^2 + F_{23}^2) & g_2 g_3 F_{23}^2 \\
		0 & g_2 g_3 F_{23}^2 & g_3^2 F_{23}^2
	\end{pmatrix}.
	\end{equation}
Diagonalization is performed first with an $ \ord{1} $ rotation between generations 1 and 2, then with an $ \ord{1} $ rotation of 2 and 3. After these, the remaining off-diagonal entries are removed with perturbative diagonalization. The result of this procedure is the rotation matrix
	\begin{equation}
	O_W = \begin{pmatrix}
	\frac{g_1}{g_{12}} & \eminus \frac{g_1 g_2^2}{g_{12} g_{123}^2} & \frac{g_2 g_3}{g_{123}^2} \\
	\eminus \frac{g_2}{g_{12}} & \eminus \frac{g_1^2 g_2}{g_{12} g_{123}^2} &  \frac{g_3 g_1}{g_{123}^2} \\
	\frac{g_2^2 g_3}{g_{12}^3 } \frac{F_{23}^2}{F_{12}^2}  & \frac{g_{12} g_3}{g_{123}^2} & \frac{g_1 g_2}{g_{123}^2}
	\end{pmatrix}\label{eq:Wmixing}, \qquad 
	\begin{dcases}
	g^2_{12} &= g_1^2 + g_2^2 \\ 
	g^4_{123} &= g_1^2 g_2^2 + g_2^2 g_3^2 + g_3^2 g_1^2  
	\end{dcases},
	\end{equation}
where we have kept only the LO term in each entry. The resulting masses are 
	\begin{equation}
	\tilde{M}_W^{2} = O_W\transpose M_W^2 O_W = \diag\big(g_{12}^2 F_{12}^2,\; g_{123}^2 F_{23}^2, 0)
	\sim  m_1^2 \, \diag( \varepsilon^2,\; \varepsilon^4, 0)  ~.
 	\end{equation}

We are now ready to analyze the stability of the mass term of the light Higgs field under radiative corrections generated by its 
interaction with gauge and link fields.  Rather than analyzing the whole series of multi-loop corrections in a systematic way, it is convenient to start analyzing first the 2--3 sector in the limit of two sites only. We will work in the flavor basis, i.e. considering the fields $H_2$ and $H_3$, rather than the mass eigenstates $\tilde{H}_2$ and $\tilde{H}_3$, which differ from the former by a small mixing angle of 
$\ord{\varepsilon^2}$, as shown in Eq.~(\ref{eq:OH}). 

We begin by considering the two-site limit before generalizing the discussion to the full model. After the diagonal symmetry breaking induced by the link fields, we are left with a massless gauge field,
$\tilde{W}_{3\mu}$,  and a massive field, $\tilde{W}_{2\mu}$, with mass $\tilde{M}_{W_2} \sim g F_{23} \sim \varepsilon m_2$. 
The 1-loop diagrams involving the heavy $\tilde{W}_2$ do not destabilize $m_3$, as can be understood by na\"ive dimensional
considerations on the following topologies:
\begin{align}
	\ineqgraphics{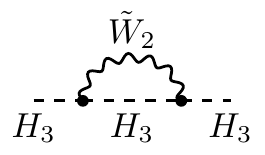} \quad + \quad 
\ineqgraphics{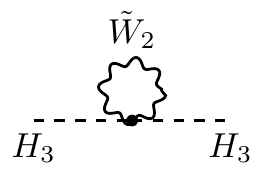}  
	 & ~\longrightarrow~~ \delta m_3^2 \sim  \frac{1}{16\pi^2}  g^2 m_{A_2'}^2 \sim \varepsilon^4 m_2^2   \sim m_3^2,
\label{eq:higgs_mass_gauge_1}
\end{align}
again with $\varepsilon = (4\pi)^{\eminus 1}$.
At the 2-loop order, we can consider diagrams of the type
\begin{align}
	\ineqgraphics{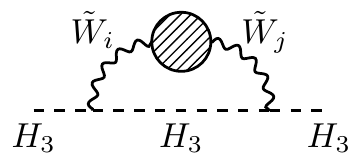}, \label{eq:2_loop_gauge}
\end{align}
where
\begin{align}
	\ineqgraphics{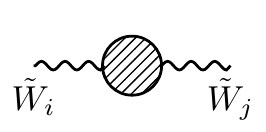}\; = ~~ \ineqgraphics{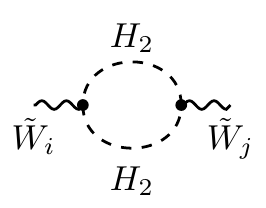}\; +  \; \ineqgraphics{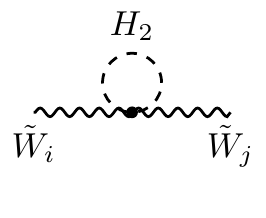}.
\end{align}
Although the 2-loop connection to the scale $m_2$ is already safe from our point of view (the two loops generate sufficient suppression), it is worth noticing that, $m_2$ being the heaviest mass in the diagram, we can in a first approximation ignore the vector boson masses in the propagators. One can then show that these leading terms cancel out completely when summing over all the possible gauge boson configurations.\footnote{This is readily seen in the gauge-basis for the vector bosons, where $ W_3 $ couples exclusively to $ H_3 $ and $ W_2 $ likewise to $ H_2 $.} This is important when we consider the extension to three sites.

The trilinear coupling $\kappa_{23} F_{23} H_2^\dagger \Omega_{23} H_3$, being proportional to $F_{23}$, leads to a ``soft" correction, not proportional to heavier scales:
\begin{align}
	\ineqgraphics{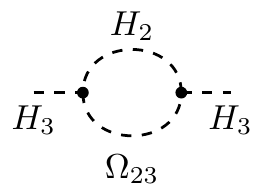}  & \longrightarrow~ \delta m_3^2 \sim \frac{1}{16\pi^2} \kappa_{23}^2 F^2_{23} \sim \varepsilon^4 m_2^2 \sim m_3^2~.
	\label{eq:H3H2}
\end{align}
Note that the appearance of $m_2$ in (\ref{eq:H3H2}) is purely conventional: the dimensionful coupling in the 
trilinear vertex implies only a logarithmic dependence from the heavy scale.

A potential problem arises with
\begin{align}\label{eq:Homegaloop}
	\ineqgraphics{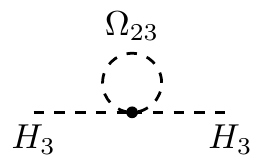}   & \longrightarrow~ \delta m_3^2 \sim  \frac{1}{16\pi^2}\, \lambda_3^-\, m^2_{\Omega_{23}} ~,
\end{align}
which might be  dangerous depending on the value of $m^2_{\Omega_{23}}$, which we have not addressed yet. 
To this end we consider the leading 1-loop correction to the $\Omega_{23}$ mass
\begin{align}\label{eq:omegaWloop}
	\ineqgraphics{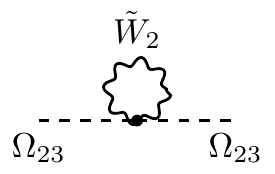}   & \longrightarrow~ \delta m^2_{\Omega_{23}}   \sim \frac{1}{16\pi^2} g^4 F^2_{23}
	\sim m_3^2~,
\end{align}
from which we conclude that the two-site case is safe for what concerns corrections from the gauge--link sector. 

We have now collected all the ingredients necessary to discuss the stability in the three-site case. Compared to the 
 two-site case, the following generalizations can be made:
\begin {itemize}
\item In the three-site case there is 
a direct coupling of the light Higgs to the heavy $\tilde{W}_1$, i.e. the analog of the diagrams in (\ref{eq:higgs_mass_gauge_1}) 
with $\tilde{W}_2 \to \tilde{W}_1$.
However, as we see from the gauge boson mixing matrix \eqref{eq:Wmixing}, this coupling 
is suppressed by $(F_{23}^2/{F_{12}^2})^2 \sim \varepsilon^8$,
which is enough to render the contribution safe.
\item
Concerning the generalization of the diagram \eqref{eq:2_loop_gauge}, we first observe that diagrams involving one or two $\tilde{W}_1$ are suppressed again by one or two factors of $(F_{23}^2/{F_{12}^2})$. Moreover, we recall the argument about the 
cancellation of the leading contributions  in the limit of vanishing gauge boson masses. The same reasoning also holds upon exchanging $ H_2 $ in the loop with $ \Omega_{12} $ or $ H_1 $.
These two arguments are sufficient to show that this class of diagrams does not spoil the hierarchy either. 
\item The 2-loop diagrams with scalars only connecting the scale $m_3$ with $m_1$, such as
\begin{align}
	\ineqgraphics{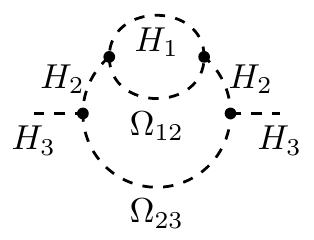}
\end{align}
benefit from the suppression induced by the coupling, similarly to what happens for the diagram in (\ref{eq:H3H2}).
This makes them effectively 4-loop (from the point of view of the $ \varepsilon$ counting), 
 hence sufficiently suppressed contributions.

\item{}
The only diagram that becomes more problematic in the three-site case is the diagram \eqref{eq:omegaWloop},
with  $\tilde{W}_2 \to \tilde{W}_1$, which results in a correction $\delta m^2_{\Omega_{23}} \sim  m_2^2$. This affect the Higgs mass when nested into diagram~\eqref{eq:Homegaloop} and would spoil the stability unless the following condition is satisfied:
\begin{equation}
\label{eq:cond_I}
\mathbf{I)}  \qquad  \lambda^-_{3} = \cO(\epsilon^2)~. 
\end{equation}
This condition is exactly the same condition we derived at the tree-level in order to ensure $m_3 \sim v_{\textrm{EW}} \ll F_{23}$. 
\end{itemize}

\section{Realistic model}
\label{sec:full_model}

Having investigated the stability of the gauge-link sector in the simplified model, we are ready to extend the fermion 
sector of the model and consider an extended setup able to provide a realistic description of fermion mixing. For this purpose, 
we introduce a vector-like $ \SU(2) $ doublet for each site, such that the matter content of the model is the content summarized in
Tab.~\ref{tab:min_eff_model}. In terms of an extra-dimensional UV completion, these states can be identified with the ubiquitous 
Kaluza-Klein (KK) excitations associated with the quark doublets. 

The fermionic part of the Lagrangian takes the form 
	\begin{multline} \label{eq:L_3-site_fermion}
	\L \supset -  \overline{Q} M Q - \bigg( 
	\overline{Q} (\hat{y}^u_\alpha H^{\ast}_\alpha) u_\RR + \overline{q}_\LL (y^u_\alpha H^{\ast}_\alpha) u_\RR + \overline{Q} (\hat{y}^d_\alpha H_\alpha ) d_\RR + \overline{q}_\LL (y^d_\alpha H_\alpha) d_\RR \\
	+ \overline{Q}_L (x_a \Omega_a) Q_R + \overline{q}_L (\hat{x}_a \Omega_a) Q_R \hc \bigg),
	\end{multline} 
where the fermions fields are to be understood as 3-component vectors in generation space, and $ M = \diag(M_1,M_2,M_3) $. The $3\times 3$ matrices of scalars and Yukawa couplings are given by (an identical expression holds for the $ \hat{y}^{u,d}_\alpha $)
	\begin{equation}
	y^{u,d}_\alpha H_\alpha = \diag\big(y^{u,d}_1 H_1, \; y^{u,d}_2 H_2, \; y^{u,d}_3 H_3\big)~,
	\end{equation}
and 
	\begin{equation}
	x_a \Omega_a = 
	\begin{pmatrix}
		0 & x_{12} \Omega_{12} & 0 \\
		x_{12} \Omega^\dagger_{12} & 0 & x_{23} \Omega_{23} \\
		0 & x_{23} \Omega^\dagger_{23} & 0
	\end{pmatrix}~,
	\qquad \hat{x}_a \Omega_a = 
	\begin{pmatrix}
		0 & \hat{x}_{12} \Omega_{12} & 0 \\
		\hat{x}_{21} \Omega^\dagger_{12} & 0 & \hat{x}_{23} \Omega_{23} \\
		0 & \hat{x}_{32} \Omega^\dagger_{23} & 0
	\end{pmatrix}~. \label{eq:x_aOmega_a}
	\end{equation}
We have let $\hat{x}_{ij} \neq \hat{x}_{ji}$ and $x_{ij} = x_{ji}$ to reflect the relations that holds for couplings originating from 5D profiles of the fermionic fields (for complex couplings we would have $x_{ij} = x_{ji}^*$ instead).

Mirroring the mass hierarchy of the scalar sector, the fermion mass terms are taken to be hierarchical with the heavy fields belonging to the first site. Here too, it is the role of the link fields to generate the off-diagonal mass terms
	\be\label{eq:Mij}
	M_{ij}= x_{ij} F_{ij} \qquad  \textrm{and}  \qquad  \hat M_{ij} = \hat x_{ij} F_{ij}
	\ee
upon receiving VEVs. We find that a suitable ordering of the fermion mass terms that reproduce the flavor structure of the SM is given by 
	\begin{equation}
	\label{eq:MHF}
	M_1 \sim m_1  \gg \; M_2 \sim \varepsilon\, m_1 \; \gg \; M_{12},\  \hat M_{12}
	\sim \varepsilon^2\, m_1 \; \gg \; M_3 \sim \varepsilon^3\, m_1	
	\; \gg \; M_{23},\  \hat M_{23} \sim \varepsilon^4\, m_1~,
	\end{equation}
which can easily be accommodated setting the adimensional couplings $x_{ij}$ and $\hat x_{ij}$ of $\ord{\varepsilon}$. This is 
perfectly consistent with the hierarchy of the scalar masses assumed in Eq.~\eqref{eq:scaleratios}.  Note, however,  that in the fermion 
sector  we interchange the relative ordering between off-diagonal and diagonal mass terms: as we shall clarify below,
this condition is necessary in order to avoid a too large mixing between the $ Q_\LL $ and $ q_\LL $ states.

\begin{table}
	\centering
	\begin{tabular}{| c | c c|}
	\hline\hline
	Fields & $ \SO(1,3) $ & $ [\SU(2)]^3 $ \\
	\hline
	$ q_i $ & $ (\tfrac{1}{2},\, 0) $ & $ \rep{2}_i $ \\
	$ Q_i $ & $ (\tfrac{1}{2},\, 0) \oplus (0,\, \tfrac{1}{2}) $ & $ \rep{2}_i $ \\
	$ u_i $ & $ (0,\, \tfrac{1}{2}) $ & $ \rep{1} $ \\
	$ d_i $ & $ (0,\, \tfrac{1}{2}) $ & $ \rep{1} $ \\
	\hline
	$ H_i $ & 0 & $ \rep{2}_i $  \\
	$ \Omega_{i(i+1)} $ & 0 & $ \rep{2}_i \otimes \repbar{2}_{i+1} $ \\
	\hline \hline
	\end{tabular}
	\caption{Matter field content of the full model.}
	\label{tab:min_eff_model}
\end{table}

\subsection{Flavor structure}
With the introduction of the vector-like doublets $ Q $, we open up for intergenerational mixing of the fermions via the link fields. 
We shall not endeavor to make a precision study of the resulting rotation matrices, as we are still working with a toy model. Rather, we wish to illustrate that the hierarchical texture of the  Cabibbo-Kobayashi-Maskawa (CKM) matrix follows directly from the assumption of a hierarchy in the mass terms. As such, the texture 
can be expected to be a generic feature in all similar models.

We begin by determining the mass states of the left-handed quark doublets prior to EWSB. 
After the link fields acquire VEVs, $ \langle \Omega_a \rangle = F_a $, the $6 \times 3$ mass term for the fermion doublets is of the form  
\begin{equation}
	\L \supset - \big( \overline{q}_\LL \; \; \overline{Q}_\LL 	\big)
	M_Q	Q_\RR , \qquad  M_Q = \binom{\hat{x}_a F_a }{M + x_a F_a}~. 
\end{equation}
$ M_Q  $ has rank 3, ensuring three massless left-handed doublets, which are identified as SM fermions, while the massive states decouple from physics at the EW scale.
The mass eigenstates of the fields are determined from the singular value decomposition of the mass matrix. Adopting the notation defined in (\ref{eq:Mij}), the rotation matrix $ U_\LL $, such that $ U_\LL\transpose M_Q\, M_Q\transpose  U_\LL $ is diagonal, is found to be\footnote{ The degeneracy between the lightest massive and the three massless states is lifted at $ \ord{\varepsilon^6} $. The eigenvalues and leading eigenprojections, thus has to be determined to $ \ord{\varepsilon^{12}} $ to achieve $ \ord{\varepsilon^6} $ accuracy for $ U_\LL $.} 
\begin{equation}
	U_\LL =
	\begin{pmatrix}
 		1 & 2 \frac{ \hat{M}_{12} \hat{M}_ {23} M_{23}}{M_ 2 M_ 3^2} & -2 \frac{ \hat{M}_{12} \hat{M}_{32} }{M_ 2^2} & \frac{\hat{M}_{12} M_{12} }{M_ 1^2} & \frac{\hat{M}_{12} }{M_2} & -\frac{\hat{M}_{12} M_{23} }{M_2 M_3} \\
		-\frac{\hat{M}_{12} \hat{M}_{23} M_{23} }{M_2 M_3^2} & 1 & 2\frac{ \hat{M}_{23} \hat{M}_{32} M_{23} }{M_ 2 M_ 3^2} & \frac{\hat{M}_{21} }{M_1} & -\frac{\hat{M}_{21} M_{12} }{M_1 M_2} & \frac{\hat{M}_{23} }{M_3} \\
		\frac{\hat{M}_{12} \hat{M}_{32} }{M_ 2^2} & -\frac{\hat{M}_{23} \hat{M}_{32} M_{23} }{M_2 M_3^2} & 1 & \frac{M_{12} \hat{M}_{32}  }{M_1^2} & \frac{\hat{M}_{32} }{M_2} & -\frac{\hat{M}_{32} M_{23} }{M_2 M_3} \\
		\frac{\hat{M}_{12} M_{12} }{M_1 M_2} & -\frac{\hat{M}_{21} }{M_1} & \frac{M_{12} \hat{M}_{32} }{M_1 M_2} & 1 & -\frac{M_{12} }{M_1} & -\frac{\hat{M}_{21} \hat{M}_{23} }{M_1 M_3} \\
		-\frac{\hat{M}_{12} }{M_2} & \frac{\hat{M}_{21} M_{12} }{M_1 M_2} & -\frac{\hat{M}_{32} }{M_2} & \frac{M_{12} }{M_1} & 1 & \frac{\hat{M}_{12}^2 M_{23}}{M_2^2 M_3}-\frac{M_{23}}{M_ 2}  \\
		\frac{\hat{M}_{12} M_{23} }{M_2 M_3} & -\frac{\hat{M}_{23} }{M_3} & \frac{\hat{M}_{32} M_{23} }{M_2 M_3} & \frac{M_{12} M_{23} }{M_1^2} & \frac{M_{23} 	}{M_2} & 1
	\end{pmatrix},
	\label{eq:ULeps}
\end{equation}
	where, in order to keep the expression manageable, we show only the leading contribution in the $\varepsilon$ expansion
	for each entry. The left-handed doublet mass eigen-states are then given by 
	\begin{equation}
	\binom{\tilde{q}_\LL}{\tilde{Q}_\LL} = U_\LL\transpose \binom{q_\LL}{Q_\LL}~,
	\end{equation} 
and the parametric scaling of the different elements of $U_\LL$ in powers of $\varepsilon$  is
\begin{equation}
	U_\LL \sim
	\begin{pmatrix}
		1 & \varepsilon^3 & \varepsilon^4 & \varepsilon^4 & \varepsilon & \varepsilon^2 \\
		\varepsilon^3 & 1 & \varepsilon^5 & \varepsilon^2 & \varepsilon^3 & \varepsilon \\
		\varepsilon^4 & \varepsilon^5 & 1 & \varepsilon^6 & \varepsilon^3 & \varepsilon^4 \\
		\varepsilon^3 & \varepsilon^2 & \varepsilon^5 & 1 & \varepsilon^2 & \varepsilon^3 \\
		\varepsilon & \varepsilon^3 & \varepsilon^3 & \varepsilon^2 & 1 & \varepsilon^3 \\
		\varepsilon^2 & \varepsilon & \varepsilon^4 & \varepsilon^6 & \varepsilon^3 & 1
	\end{pmatrix}~.
\end{equation}

The mixing between the $ q_\LL $ and $ Q_\LL $ fields is responsible for populating the off-diagonal entries of the Yukawa coupling 
between the massless left-handed doublets and the right-handed singlets, $ u_\RR $ and $ d_\RR $. In terms of the mass eigenstates before EWSB, the Yukawa coupling for the up-type quarks reads
	\begin{equation}
	\L \supset - \big( \overline{\tilde{q}}_\LL \;\; \overline{\tilde{Q}}_\LL \big) U_\LL\transpose 
	\binom{
	y_\alpha^u H^\ast_\alpha }{ \hat{y}_\alpha^u H^\ast_\alpha}	u_\RR .
	\end{equation}
At the EW scale the heavy $ \tilde{Q} $ fields decouple and the resulting effecctive Yukawa coupling for the light (SM-like) 
fermions is contained in the upper $ 3 \times 3 $ matrix. After EWSB, taking the hierarchy~\eqref{eq:higgsVEVs} of the Higgs VEVs, namely  $v_1 \sim \varepsilon^4 v_3$ and $v_2 \sim \varepsilon^2 v_3$, 
and the structure~\eqref{eq:ULeps} of  $U_\LL$ into account, leads to the mass matrix 
\begin{equation} \label{eq:up-quark_mass_matrix}
	m_u = 
	\begin{pmatrix}
		 y_1^u v_ 1 & -\frac{\hat{M}_{12} }{M_2} \hat{y}_2^u v_ 2 & \frac{\hat{M}_{12} M_{23} }{M_2 M_3} \hat{y}_3^u v_3 \\
		-\frac{\hat{M}_{21} }{M_1} \hat{y}_1^u  v_1  & y_2^u v_2  & -\frac{\hat{M}_{23}   }{M_3} v_ 3 \hat{y}_3^u \\
		\ord{\varepsilon^7} & -\frac{\hat{M}_{32} }{M_2} \hat{y}_2^u v_2  & y_3^u v_3  
	\end{pmatrix}
	\sim
	\begin{pmatrix}
		\varepsilon^4 & \varepsilon^3 & \varepsilon^2 \\
		\varepsilon^6 & \varepsilon^2 & \varepsilon \\
		\ord{\varepsilon^7}  & \varepsilon^5 & 1
	\end{pmatrix} \times v_{\textrm{EW}}
\end{equation}
of the up-type quarks.
The rotation $ V^u_\LL $ required to bring the left-handed $ u $ quarks to the mass basis can then be determined by singular value decomposition of $ m_u$, or by the diagonalization of $m_u m_u\transpose $.
The latter can be done perturbatively in  $\varepsilon$: given the level of approximations employed so far, we can compute 
the eigenvalues of $m_u m_u\transpose$ up to $\ord{\varepsilon^6}$. Since the degeneracy of the two lightest squared masses is lifted only at order $ \varepsilon^4 $,  this allows for determining $ V^u_\LL $ up to $\ord{\varepsilon^2}$, which is the level of accuracy 
necessary for describing the structure of the CKM matrix. 

The determination of $ m_d $ and the left-handed rotation matrix $ V_\LL^d $  for the down quarks is entirely analogous 
to that in the up sector. In fact, since the structure of both $ m_u $ and $ m_d $ is determined by the $ U_\LL $ rotation, the mass matrix~\eqref{eq:up-quark_mass_matrix} is valid for $ m_d $ upon substitution 
$ y^u (\hat y^u) \to y^{d} (\hat y^d)$. 
One thing to bear in mind is that, within the SM, down-type quark masses  exhibit a different hierarchy with respect to up-type ones. 
This feature can be implemented in the model with the minor adjustment $ y_2^d \sim \hat{y}_2^d \sim y_3^d \sim \hat{y}_3^d \sim \varepsilon $. The result of this procedure is the physical CKM matrix 
\begin{equation}
\label{eq:CKM}
	V_\mathrm{CKM} = V_L^{u^\dagger}V_L^d = 
	\begin{pmatrix}
		1 & V_{12}  & V_{13}  \\
		-V_{12} & 1 & V_{23}  \\
		V_{31}   & -V_{23}   & 1 \\
\end{pmatrix}
\sim
\begin{pmatrix}
	1 & \varepsilon & \varepsilon^2 \\
	\varepsilon & 1 & \varepsilon \\
	\varepsilon^2 & \varepsilon & 1
\end{pmatrix},
\end{equation}
where
\begin{equation}
	\begin{split}
		V_{12} &= \frac{\hat{M}_{12}}{M_2} \Delta y_2 , \qquad
		V_{23} = \frac{\hat{M}_{23}}{M_3}\Delta y_3 , \\
		V_{13} &= \frac{\hat{M}_{12}}{M_2}\left(\frac{\hat{M}_{23}}{M_3}\frac{\hat{y}_2^u}{y_2^u} - \frac{M_{23}}{M_3} \right)\Delta y_3 , \\
		V_{31} &= \frac{\hat{M}_{12}}{M_2}\left(\frac{M_{23}}{M_3} - \frac{\hat{M}_{23}}{M_3}\frac{\hat{y}_2^d}{y_2^d} \right)\Delta y_3 . 
	\end{split}
\end{equation}
and
\begin{equation}
\Delta y_i = \frac{\hat{y}_i^u}{y_i^u} - \frac{\hat{y}_i^d}{y_i^d} .
\end{equation}

With $\varepsilon =  (4\pi)^{\eminus 1}$ and $\ord{1}$ parameters in the range $[\frac{1}{3},3]$, 
the parametric structure in \eqref{eq:CKM} fits well with the observed hierarchical 
structure of the CKM matrix~\cite{PDG}. Combining this  fact with the compatibility of the eigenvalues 
of both $m_u$ and $m_d$ with observations, we conclude that this framework provides a realistic 
description of the SM quark Yukawa couplings. Moreover, the construction is such that 
both $m_u$ and $m_d$  are diagonalized by tiny right-handed rotations. 
To a large extent, the model provides an explicit 
dynamical realization of the minimally-broken $\U(2)^3$ quark-flavor symmetry, which is known to provide a good 
description of the SM spectrum avoiding dangerous new sources of flavor violation in the dimension-six operators
obtained by integrating out heavy fields at the TeV scale~\cite{Barbieri:2011ci,Barbieri:2012uh}.

\subsection{Stability}
Having verified that the postulated hierarchy of scales provides a realistic description of the SM spectrum, we proceed to 
investigate the stability of the Higgs sector in this more complete setup. Building on the discussion of Section~\ref{sec:gauge_stability}, we carefully consider the impact of the vector-like fermions and the corresponding intergenerational couplings. 
Once again, our primary concern is to ensure the stability of $ m_3 $, which is the lightest scale of the model. 

On general grounds, the largest ratio of mass parameters in the model is $ m_1^2/m_3^2 \sim \varepsilon^{\eminus 8} $. 
Hence the $ m_1^2 $ contribution to $ \delta m_3^2 $ needs to be suppressed by 4 loop factors (or, equivalently, small couplings), in order  
to avoid an unreasonably large tuning. A systematic search for diagrams with potentially problematic contributions needs to extend up to the 3-loop order. What protects the hierarchy of scales is the largely decoupled nature of the fields living on the different sites. Any interaction between them necessarily involve the link fields. The link between $H_3 $ and $ H_1 $ must therefore be mediated through two link fields and some field living on the second site. Any diagram complicated enough to contain all of these fields necessarily appear at the 3-loop order or higher, thereby protecting 
$m_3^2 $ against large corrections. 

	\begin{figure}
	\centering
	\includegraphics[width=.7\textwidth]{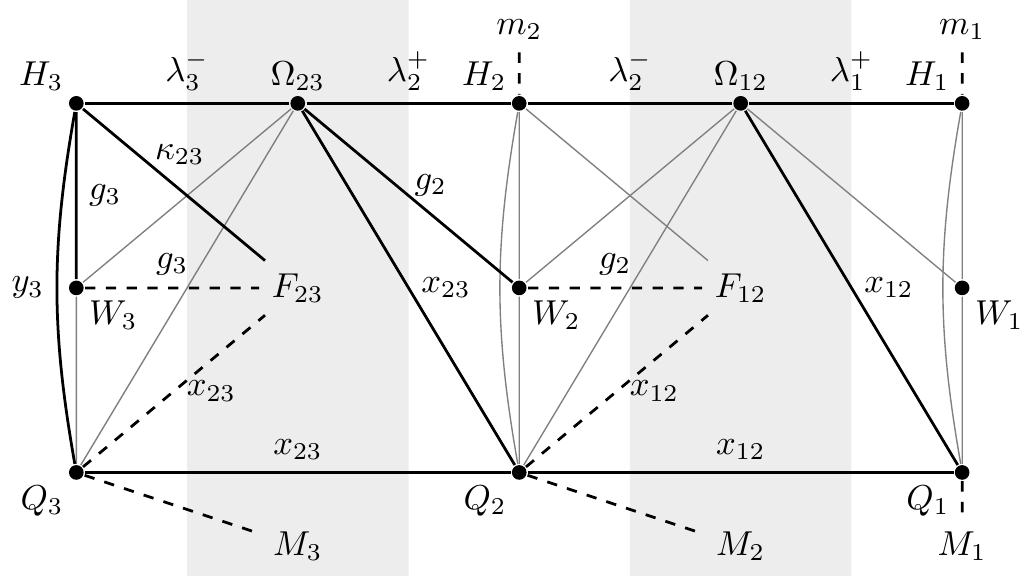}
	\caption{Diagram showing how to connect $ H_3 $ (top left) to the various mass parameters of the theory 
	at a given order in the expansion in terms of loops (in one-to-one correspondence with the lines) and couplings 
	(indicated over each line), within the full model (see main text). 
	The faint gray lines correspond to loops/couplings which are less dangerous for the stability of the hierarchy. 
	The vertical bands groups fields and mass parameters associated to different energy scales.}
	\label{fig:interaction_diagram}
	\end{figure}

To make our reasoning more systematic, we consider the known topological relations between the loop order of a diagram and the 
couplings this diagram contains. A mass correction diagram to $ H_3 $ has two external legs, and the loop order $ \ell $ is given by 
	\begin{equation} \label{eq:loop_count}
	2 \ell = 2 n_\lambda + n_g + n_x + n_y + 2n_\kappa - n_F~,
	\end{equation}
where $ n_\lambda $ is the number of $ \lambda_i^{\pm} $ couplings; $ n_g $ denotes the number of gauge couplings ($ g_i $); 
$ n_y $ and $ n_x $ indicate the number of Yukawa couplings $y_i (\hat y_i)$ and $x_{ij} (\hat x_{ij})$, respectively; 
$n_\kappa $ is the number of trilinear scalar couplings ($ \kappa_{ij} $); and $ n_F $ denotes the number of VEV insertions ($ F_{ij} $).
The possible couplings among fields are limited: 
the ways in which the different fields couple to one another is summarized  by the diagram in Fig.~\ref{fig:interaction_diagram}.   

The diagram in Fig.~\ref{fig:interaction_diagram}  provides an effective tool for a systematic analysis of the stability of $m_3$.
To do so, we need to begin and end at $H_3$ in the top left corner, corresponding to 
considering a  two-point diagram with two external $H_3$ lines, and connect to some other mass scale. 
For instance, we can connect  $H_3$ to  $\Omega_{23}$ via  the $ \lambda_3^{-} $ coupling, and from there to the $ H_2 $ field 
via $ \lambda^{+}_{2} $. At this point, we can get an $ m_2^2 $ mass-insertion, generating a correction to $m^2_3$ proportional to $m^2_2$.
This is the shortest way to go from $ H_3 $ to $ m_2 $: any other path would involve more couplings and, correspondingly, a stronger 
loop suppression. From this minimal path we deduce that $m_3^2 $ receives a correction from $ m_2^2 $ at the 2-loop order, 
with coefficient $ \lambda_3^{+} \lambda_2^- $. A point of caution: the figure does not guarantee the existence of a diagram, but it can easily be checked whether one exists given the couplings involved. In general, $ H_3 $ can be connected to any of the masses and VEVs in the figure, providing the end points of any line, picking up all the couplings along the way.\footnote{The quartic couplings should  be counted only once in the paths back and forth.}  
Then one can trace a path back to $ H_3 $ (typically the same path) to connect to the second $ H_3 $ leg of a mass correction diagram. From Eq.~\eqref{eq:loop_count}, it follows that every one of the solid lines in the figure increases the resulting Feynman diagram by 1-loop order (half going out and half coming back again). 

From the inspection of the diagram in Fig.~\ref{fig:interaction_diagram} we identify three potentially problematic contributions to 
$ \delta m_3^2 $, i.e.~contributions where the loop suppression alone is not enough to ensure the smallness of the corrections. 
These corresponds to the paths connecting $ H_3 $ to $ F_{12} $ or $ M_2 $ at the 2-loop level. 
The possibility of coupling $ H_3 $ to $ W_2 $ and then $ F_{12} $ through $ \lambda_3^{-} $ and $ g_2 $ was already identified in Section~\ref{sec:gauge_stability}: the resulting contribution is
	\begin{equation}
	\delta m_3^2 \sim \dfrac{1}{(4\pi)^4} \lambda_3^- g_2^4 F_{12} \sim \dfrac{\lambda_3^-}{\varepsilon^2} m_3^2~ 
	\end{equation}
and can ultimately be tamed by requiring a small $\lambda_3^-$, cf. condition \eqref{eq:cond_I}.
The seemingly problematic contributions induced by the heavy fermions arise from coupling $ H_3 $ to $ Q_2 $.
Following the path via $y_3 (\hat y_3)$  and $ x_{23} (\hat x_{23})$, the mass insertions on the $ Q_2 $ line 
implies   
	\begin{equation}
	\ineqgraphics{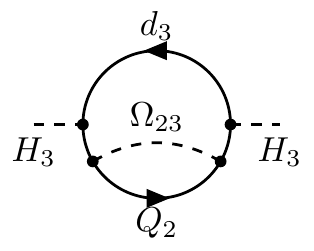} \longrightarrow~
\delta m_3^2 \sim  \dfrac{1}{(4\pi)^4} \hat{y}_3^2 x_{23}^2 M_2^2 \sim \dfrac{x_{23}^2}{\varepsilon^2} m_3^2~.   
	\end{equation} 
It is easy to realize that this is not a problematic contribution once we assume 
\begin{equation}
\label{eq:cond_II}
\mathbf{II)} \qquad
x_{ij} = \ord{\epsilon}~,  
\end{equation}
which is required already at the tree level in order to ensure the hierarchy in Eq.~\eqref{eq:MHF}.
Incidentally, we note that having somewhat suppressed intergenerational couplings is easily reconciled with the 5D picture, where such couplings arise from overlap integrals. The second 2-loop path from $ H_3 $ to $ Q_2 $ on Fig.~\ref{fig:interaction_diagram} is suppressed by both $ \lambda_3^- $ and $ x_{23} (\hat x_{23})$ and  therefore sub-dominant.
 
By inspection of the other paths in Fig.~\ref{fig:interaction_diagram}, one quickly realizes 
that there are no other problematic radiative contributions to $ m_3 $. Furthermore, we note that  
the next-to-lightest mass, $ M_3 $, is protected from $ M_2 $ by the smallness of $ x_{23}$ ($\hat x_{23}$) and by loop factors, 
ensuring its stability. We thus conclude that if the two conditions in Eq.~\eqref{eq:cond_I} and \eqref{eq:cond_II}
are satisfied, the mass hierarchy of the full model is stable under radiative corrections.

\section{Conclusions}

Models addressing the hierarchy problem via flavor-blind dynamics are in strong tension with high-$p_T$ experiments. 
On the other hand, signals of flavor-non-universal interactions seem to emerge from 
low-energy experiments~\cite{deSimone:2020kwi}. 
Motivated by these observations, we have analyzed the stability of the Higgs sector of a model addressing the flavor problem via a hierarchical spectrum of scalar, fermion, and flavor-non-universal massive gauge fields.
The spectrum of new states spans three orders of magnitude:
from the TeV scale up to about  $10^3$~TeV. 

The model we have considered, which is only a toy model, provides a representative example of more realistic 
constructions based on the idea that the generation index refers to a specific localization 
in an extra space-like dimension. As we have shown, this construction provides an a posteriori explanation of the observed 
hierarchies in the SM fermion spectrum in terms of the scale hierarchies in the underlying theory: the light generations are light
not because of small couplings, but because their effective coupling to the Higgs field is mediated by the exchange of heavy 
new states. A similar phenomenon occurs for the smallness of the off-diagonal entries of the CKM matrix.

In such a context, one might be worried about the impact of the heavy states in destabilizing the SM ground state. 
Our analysis shows that  the effective mass term 
of the SM-like Higgs field does not receive large corrections proportional to the heavy fields.
More precisely, we have shown that the Higgs mass receives at most relative corrections of $\ord{1}$, 
provided the two conditions in Eqs.~\eqref{eq:cond_I} and~\eqref{eq:cond_II} are satisfied. The condition~\eqref{eq:cond_II} is the natural requirement of 
a not too-large mixing between SM-like and heavy vector-like fermions. Condition~\eqref{eq:cond_I}, which is the only  
real tuning of the model, is nothing but a manifestation of the little hierarchy problem in this context: 
an $\ord{1\%}$ tuning that we need to impose in order to screen the electroweak scale 
from the lightest NP states of the model, which are expected to be at the TeV scale. 
This latter problem could be addressed in more sophisticated versions of the model, with additional 
protective symmetries acting on the Higgs sector; however, this is beyond of the scope of this paper. 

In summary, our analysis provides a concrete example of how to convert 
scale hierarchies into flavor hierarchies, in the context of models with flavor-non-universal gauge interactions, 
and the proof that this exchange does not worsen the electroweak hierarchy problem. 
Actually, this type of construction makes the electroweak hierarchy problem less severe 
compared to flavor-blind frameworks, due to the weaker bounds on TeV-scale NP 
coupled mainly to the third generation.

\subsection*{Acknowledgments}
This project has received funding from the European Research Council (ERC) under the European Union's Horizon 2020 research and innovation program under grant agreement 833280 (FLAY), and by the Swiss National Science Foundation (SNF) under contract 200021-175940.


\sectionlike{References}
\vspace{-10pt}
\bibliography{References} 
\end{document}